\documentclass{aa}  

\usepackage{graphicx}
\usepackage{txfonts}
\usepackage{amsmath}
\usepackage{nicefrac}
\usepackage{booktabs}
\usepackage[switch]{lineno}
\usepackage{soul}
\setstcolor{red}
\usepackage{natbib}
\usepackage[colorlinks,citecolor=blue,linkcolor=blue,urlcolor=cyan]{hyperref}

\usepackage{orcidlink}

\begin{document}

   \title{Exploring the capability of HH 80-81 protostellar jet to accelerate relativistic particles}\authorrunning{J. M\'endez-Gallego et al.}

   \author{J. M\'endez-Gallego\inst{1}\fnmsep\thanks{\email{jmendez@iaa.es}}
        \and R. L\'opez-Coto\inst{1} %\orcidlink{0000-0002-3882-9477}
        \and E. de O\~na Wilhelmi\inst{2} %\orcidlink{0000-0002-5401-0744}
        \and R. Fedriani\inst{1} %\orcidlink{0000-0003-4040-4934} 
        \and J. Otero-Santos\inst{1}%\orcidlink{0000-0002-4241-5875} 
        \and Y. Cant\"urk \inst{2}
        }

   \institute{Instituto de Astrofísica de Andalucía (IAA), CSIC,
              18008 Granada, Spain
         \and
             Deutsches Elektronen-Synchrotron DESY, Platanenallee 6, 15738 Zeuthen, Germany  
             }

   \date{Received XXXX; accepted XXXX}

% \abstract{}{}{}{}{} 
% 5 {} token are mandatory

%\linenumbers
 
  \abstract
  % context heading (optional)
  % {} leave it empty if necessary  
   {Protostellar jets driven by massive protostars are collimated outflows producing high-speed shocks through dense interstellar medium. Fast shocks can accelerate particles up to relativistic energies via diffusive shock acceleration, producing non-thermal emission that can originate $\gamma$-ray photons. HH 80-81 is one of the most powerful collimated protostellar jets in our galaxy, with non-thermal emission detected in radio, X-ray, and $\gamma$-ray bands. Characterize the $\gamma$-ray emission originated by the accelerated particles of the region is crucial for demonstrating the capability of protostars to accelerate cosmic rays.}
  % aims heading (mandatory)
   {Our goal is to determine the particle distribution that is producing the $\gamma$-ray spectrum of HH 80-81 in order to ascertain the leptonic/hadronic origin of the $\gamma$-ray emission. We aim at associating the high-energy emission in the region with the HH 80-81 system, characterize its spectrum, and elaborate emission models based on what we expect from the diffusive shock acceleration.}
  % methods heading (mandatory)
   {We use the 15 yr database provided by the \textit{Fermi}-LAT satellite to study the high-energy emission of the jet, spanning from $\rm 300 \, MeV$ to $\rm 100 \, GeV$. In addition, we perform a source association based on positional arguments.
   Then, we employ the \texttt{naima} and \texttt{Gamera} softwares to analyze the possible mechanisms that are producing $\gamma$-rays considering the ambient conditions. We perform a radiative fitting and study the nature of the particles behind the $\gamma$-ray emission. }
  % results heading (mandatory)
   {By analyzing all the candidates to produce the $\gamma$-ray emission that we detect, we conclude that HH 80-81 is the most probable candidate to explain the $\gamma$-ray emission in the region. The detected spectrum can be explained by both hadronic and leptonic particle components.}
  % conclusions heading (optional), leave it empty if necessary 
   {}

   \keywords{cosmic rays --
             gamma rays --
             massive young stellar objects --
             IRAS 18162-2048 --
             protostellar jets --
             HH 80-81
               }

   \maketitle
%
%-------------------------------------------------------------------

\section{Introduction}

    Massive young stellar objects (MYSOs), typically characterized by masses above $\rm 8 \, M_\odot$, have been objects of study due to their unique properties and their impact on star-forming regions. Massive protostars have higher luminosities than low-mass protostars, regulating the star-forming environments heating and ionizing the medium \citep{Tan_2014, Kolligan_18}. In addition, their powerful stellar winds and jets are essential to understand the mechanical feedback of the star-forming complexes, affecting the entire population of star-forming objects in the region.  Evolutionary models indicate that MYSOs evolve rapidly, reaching the zero-age main sequence stage while still embedded within the giant molecular clouds, where the star formation takes place in the galaxy \citep{Kahn_1974,Wolfire_1987}. This rapid evolution limits their capability to continue accreting mass. However, the detection of significantly more massive stellar objects highlights the necessity for mechanisms enabling the effective mass accretion once the star begins heating the surrounding medium. Theoretical studies approach this issue using massive accretion disks and collimated jets \citep{Hosokawa_2010, Kuiper_2010}, which are consistent with observations \citep{Beltran_2016, Fedriani_18_intro, Frost_2019, Fedriani_19_intro, Backs_2023}. 
   
   Surveys carried out by \cite{Caratti_2015}, \cite{Moscadelli_2016}, and \cite{Purser_2021} on MYSOs demonstrate a significant presence of jets associated with these sources. The outflows interact with the ambient medium, commonly producing knot shocks within the jet interior, as well as the powerful termination shocks. Charged particles crossing shocks gain energy via first-order Fermi acceleration, hereafter diffusive shock acceleration (DSA) \citep{Bell_1978, DSA_78}. This is one of the most common mechanism to accelerate cosmic rays (CRs) in the Milky Way. Consequently, jets can be associated with non-thermal emission \citep{Obonyo_19}, dominating different parts of the source spectrum. Radio frequencies are dominated by synchrotron emission, showing a negative spectral index consistent with non-thermal emission \citep[e.g.][]{Araudo_2008, Anglada_18}. On the other hand, $\gamma$ rays can trace the non-thermal emission directly originated by the interactions of relativistic particles in the region.

    Herbig-Haro (HH) objects \citep{Herbig_1951, Haro_1952} are composed of bright optical structures associated with knots created in jets originated by young stellar objects, also capable to emit from radio to infrared (IR) and even X-rays \citep[see][]{Schneider_22}. These knots are produced by shocking material interactions, reaching hundreds of $\rm km \, s^{-1}$ \citep[e.g.][]{Caratti_2009, Lopez_15, Djupvik_16, Massi_23}. When ejecta interacts with the ambient medium or with slower structures within the jet \citep{Canto_2000}, strong shocks are produced, specially the termination shocks, where the ejected material in the outer parts of the outflow directly impacts the interstellar medium (ISM). 
    Consequently, HH objects usually present a clear ionization structure, showing H$\alpha$ emission in the shocked regions and emitting lines of neutral species such as OI in the surrounding gas \citep{Frank_14, Krumholz_15}. Moreover, HH objects show bipolar structure \citep{Bally_2016, Ray_2023}, pointing to their origin from a single driving star, even though they are typically located at star-forming regions interacting with other protostellar components \citep[e.g.][]{Plunkett_2013, Lopez_2022}.
    
    In this context, IRAS 18162-2048 \citep[also known as GGD27 MM1;][]{Gyulbudaghian_1978} is a MYSO with $\rm \sim 20 \, M_\odot$ powering HH 80-81 \citep{Anez_2020}, the largest collimated jet originated by a protostar in the galaxy known so far, with $\rm \sim 10 \, pc$ of projected size \citep{Masque_2015, Bally_23}. 
    Among the several knots in the entire structure, HH 80 and HH 81 spotlight the termination shock in the southern part of the jet, whereas HH 80N is the brightest object in the northern part. Both regions are located $\rm \sim 2 - 3 \, pc$ away form the protostar. The whole system is placed at the western edge of L291, its host cloud. Although \cite{Anez_2020} and \cite{Zhang_2023} reported a distance to L291 of $\rm \sim 1200 \, pc$, we use the value of $\rm 1400 \, pc$ from \cite{Zucker_2020}, based on the reddening measurements of stars in the region whose parallaxes are recorded in Gaia DR2. At this distance, the bolometric luminosity of the protostar is $\rm \sim 1.2 \times 10^4 \, L_\odot$, consistent with an OB forming star. Additionally, IRAS 18162-2048 is part of a multiple system, accompanied by GGD27 MM2 \citep{Fernandez_2011}, which drives a less energetic molecular outflow compared to MM1. \cite{Busquet_2019} also reported a protostellar cluster near the GGD27 complex, populating the region with 23 low-mass young protostars more.
    
    Regarding high-energy astrophysics, while studying HH 80-81, \cite{Carrasco-Gonzalez_2010} detected, for the first and only time, linear polarization consistent with non-thermal synchrotron emission along a protostellar jet. Taking into account the clear negative index in the radio spectrum reported by \cite{Marti_1993}, HH 80-81 stands out as the best candidate to determine whether protostellar jets are able to accelerate particles. In addition, \cite{Rodriguez-Kamenetzky_2019} reported soft and hard X-ray emission in HH 80 and HH 81 ranging from $\rm 0.3 \, keV$ to $\rm 10 \, keV$, detecting a non-thermal component which dominates the spectrum form $\rm \sim 1 \, keV$ to the highest energies. Recently, \cite{Yan2022} reported a $\rm \sim 10 \sigma$ detection of this source in low-energy $\gamma$ rays, tracing the high-energy spectrum of the protostellar system from $\rm 100 \, MeV$ to $\rm 1 \, GeV$ with 10 years of exposure time.
    %supporting the hypothesis of non-thermal emission extended to high energies.

    The confirmation of non-thermal emission from a protostellar jet has motivated studies of MYSOs as potential Galactic cosmic rays factories \citep{Araudo_2007, Bosch-Ramon_2010, Araudo_2021}. Since then, the detection and association of $\gamma$-ray emission with protostellar jets has been pursued. The cases of the proximity of HH 219 to 4FGL J0822.8–4207 \citep{Araya_22}, the flaring episodes detected in S255 NIRS3 \citep[see][]{deOna_2023}, and the unassociated \textit{Fermi}-LAT source of 4FGL J1846.9-0227 \citep{ortega_2024} remark the importance of finding a new unexpected Galactic CR source. To answer the question of whether protostellar jets are capable to accelerate particles up to high energies, one needs to determine the type of relativistic particles populating these environments, since the composition of the CRs that we directly detect is dominated by protons. Therefore, most of the literature on this topic, including this work, focuses on detecting $\gamma$-ray emission from these sources coming from accelerated particles in the source environment, providing direct insights into the characterization, if possible, of hadronic non-thermal emission. Then, to determine the hadronic/leptonic origin of the particles, several radiative models are usually proposed based on the different known mechanisms to produce $\gamma$ rays. For protons, the most usual process is the proton-proton collision, producing $\pi^0$ mesons that rapidly decay into $\gamma$ rays. Electrons, in contrast, can produce $\gamma$ rays via relativistic Bremsstrahlung due to the interaction of high-energy electrons with the ambient medium, or through inverse Compton (IC) scattering of lower-energy photon fields.

    This paper focuses on the characterization of the $\gamma$-ray emission from the protostellar jet driven by IRAS 18162-2048, and is organized as follows: in Sect. \ref{sect:analysis}, we analyze the observational high-energy data of the proximities of HH 80-81 from the \textit{Fermi}-LAT telescope; in Sect.\ref{sect:association}, we perform an association analysis to identify the source originating the $\gamma$-ray excess; in Sect. \ref{sect:gammaprod}, we characterize the particle spectrum that may originate the $\gamma$ emission detected; and in Sect. \ref{sect:conclusion}, we summary our findings in a final conclusion.
    
    %The \textit{Fermi}-LAT instrument is a spacecraft telescope specialized in high-energy wavelengths spanning from $\rm 30 \, MeV$ to $\rm 300 \, GeV$. It was launched in 2008 and it has been collecting data from the entire celestial sphere since August 4$\rm ^{th}$ 2008, providing one of the most comprehensive databases of $\gamma$-ray astronomy. Its high capability to distinguish faint sources in a bright and complex background, also with the long exposure times, makes it ideal to study low-energy $\gamma$-ray emitters such as protostellar jets.

   %Based on LAT data, the goal of this paper is to infer the properties of the particles accelerated by HH 80-81 by studying the $\gamma$-ray emission produced when those particles interact with the surrounding medium. Of course, the leptonic/hadronic origin is fundamental to determine if our source is capable of accelerating CRs. We expect that electrons get cooler rapidly, while protons are able to travel long distances. This paper is organized as follows: in Sect. \ref{sect:analysis} we analyze the observational data from \textit{Fermi}-LAT instrument; in Sect. \ref{sect:gammaprod} we studied the particle spectrum that may originate the $\gamma$ emission detected, and in Sect. \ref{sect:conclusion} we summary our findings as a conclusion.

\section{Observations and Data Analysis}
\label{sect:analysis}

\begin{figure}
    \centering
    \includegraphics[width=75mm]{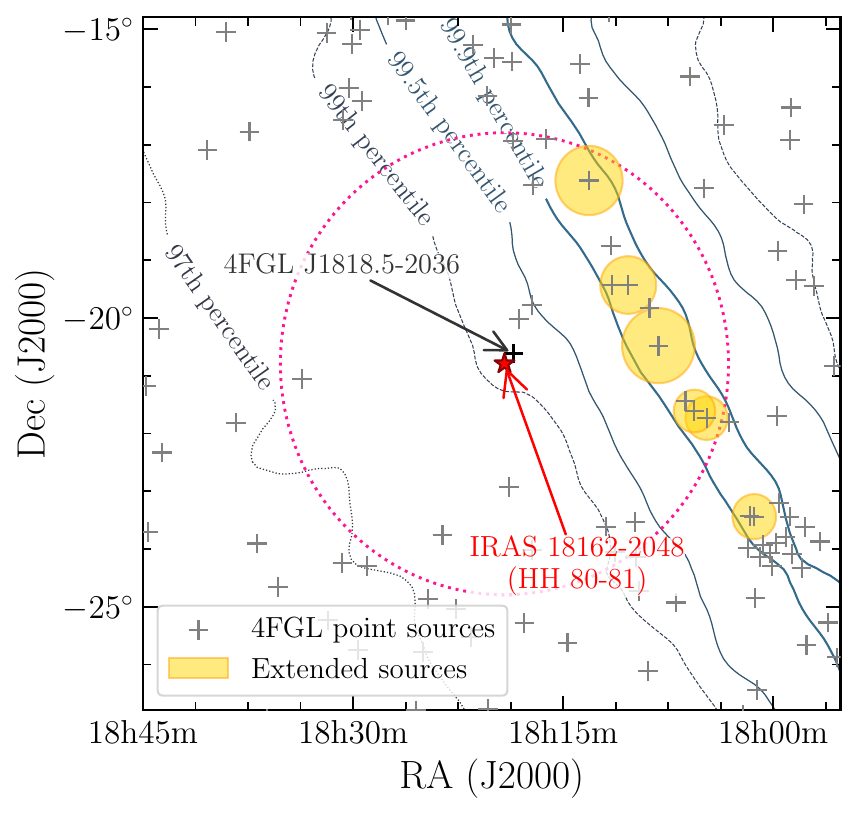}
    \caption{4FGL DR4 source map for our ROI. Contours illustrate the Galactic plane emission at the 97th, 99th, 99.7th, and 99.9th percentile based on \texttt{gll\_iem\_v07} template. Magenta circle indicates the region where the normalization parameter were set to free for computing the fitted model of the ROI. In addition, Galactic diffuse emission and isotropic emission were also set to free.}
    \label{fig:HH8081_roi}
\end{figure}

\begin{figure*}
    \centering
    \includegraphics[width=0.76\textwidth]{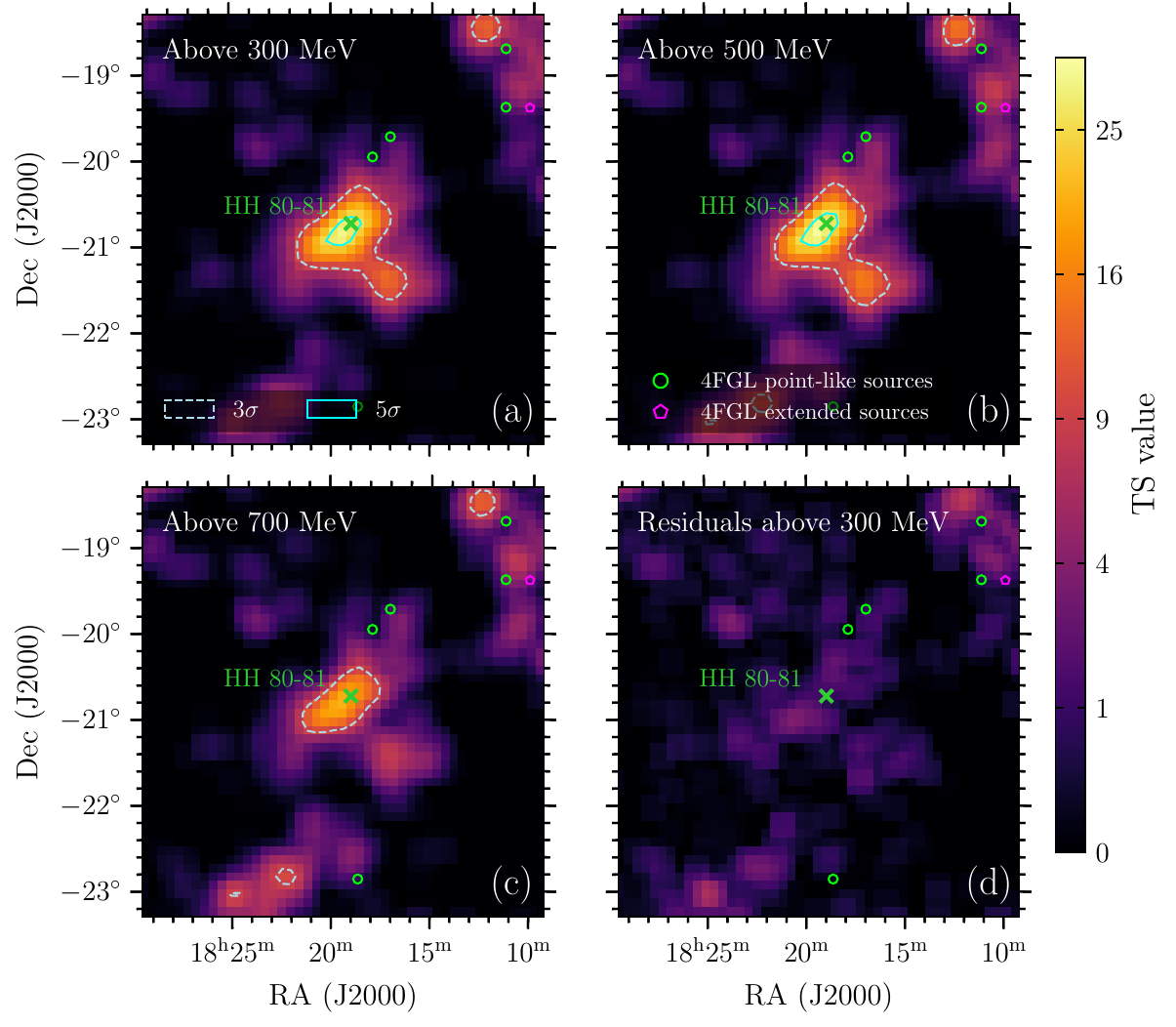}
    \caption{TS maps for a $5^\circ \times 5^\circ$ region centered in IRAS 18162-2048. Contour maps show the detection significance while color maps indicate the TS value of each spatial bin. Green cross indicates the position of the protostar driving the HH 80-81 system. Panels (a), (b), and (c) show the significance map for HH 80-81 above 300 MeV, 500 MeV, and 700 MeV respectively. Panel (d) shows the residual TS map above 300 MeV as a proof of the reliability of our fitted model.}
    \label{fig:ts_maps}
\end{figure*}

The \textit{Fermi}-LAT instrument is a spacecraft telescope specialized in high-energy wavelengths spanning from $\rm 30 \, MeV$ to $\rm 300 \, GeV$. It was launched in 2008 and it has been collecting data from the entire celestial sphere since August 4$\rm ^{th}$ 2008, providing one of the most comprehensive databases of $\gamma$-ray astronomy. Its high capability to distinguish faint sources in a bright and complex background, also with the long exposure times, makes it ideal to study low-energy $\gamma$-ray emitters such as protostellar jets.

To analyze $\gamma$-ray emission emanating from HH 80-81 proximities, we employ 15 yr of \textit{Fermi}-LAT data, ranging from August 5$\rm ^{th}$ 2008 to August 5$\rm ^{th}$ 2023. The photon data selection is performed using the standard \texttt{P8R3} (Pass 8 Release 3) \texttt{Source} class, optimal for point-like sources and long timescales \citep{Atwood_2013}. Additionally, we select \texttt{FRONT} and \texttt{BACK} events in order to  maximize the number of recorded photo-events. We employ \texttt{fermipy v1.2.2}\footnote{\href{https://fermipy.readthedocs.io/en/latest/}{https://fermipy.readthedocs.io/en/latest/}} \citep{fermipy_2017} to perform a binned likelihood analysis of our data, which is based on the \texttt{Fermi Science Tools v2.2.0}\footnote{\href{https://fermi.gsfc.nasa.gov/ssc/data/analysis/software/}{https://fermi.gsfc.nasa.gov/ssc/data/analysis/software/}} \citep{FermiTools_19}.
%The analysis of our data selection consists in a binned likelihood analysis employing \texttt{fermipy v1.2.2}\footnote{\href{https://fermipy.readthedocs.io/en/latest/}{https://fermipy.readthedocs.io/en/latest/}} \citep{fermipy_2017}, which is based on \texttt{Fermi Science Tools v2.2.0}\footnote{\href{https://fermi.gsfc.nasa.gov/ssc/data/analysis/software/}{https://fermi.gsfc.nasa.gov/ssc/data/analysis/software/}} \citep{FermiTools_19}.

The selected shape of the region of interest (ROI) is a square of $ 12^\circ \times 12^\circ$ centered at the coordinates of IRAS 18162-2048 $\rm ( RA_{J2000}=18^h 19^m 12^s.1 ,\ Dec_{J2000} = -20^\circ 47' 31'' )$ \citep{IRAS_position}. The recent \textit{Fermi}-LAT 4FGL Data Release 4 (DR4) catalog \citep{Abdollahi_2022, 4FGL-DR4_2024} is used as the base model of the ROI, including all point and extended sources within a $22^\circ \times 22^\circ$ field around the protostar (see Fig. \ref{fig:HH8081_roi}). Although the usual energy range of this type of studies spans from 100 MeV to 100 GeV, we have analyzed data from 300 MeV to 100 GeV to avoid contamination from Galactic diffuse emission at low energies (see Sect. \ref{sect:detection}). In addition, the Galactic plane diffuse emission is modeled using the most recent template \texttt{gll\_iem\_v07}, and the isotropic diffuse emission is modeled with \texttt{iso\_P8R3\_SOURCE\_V3\_v1} \footnote{\href{https://fermi.gsfc.nasa.gov/ssc/data/access/lat/BackgroundModels.html}{\small https://fermi.gsfc.nasa.gov/ssc/data/access/lat/BackgroundModels.html}}.

To compute our ROI source model, we apply a spatial binning  of $0.1 ^\circ$ per pixel, and the energy range is divided into 8 bins per decade. The selected instrument response functions (IRFs) are contained in \texttt{P8R3\_SOURCE\_V3}, being consistent with the event type that we want to analyze. We also apply a cut of events within a zenith angle above $90^\circ$ to suppress contamination from Earth Limb events. Regarding the good time interval selection, we apply the \texttt{(DATA\_QUAL>0) \& (LAT\_CONFIG==1)} filters, ensuring that the telescope was functioning correctly while our data was been taking.

Figure \ref{fig:HH8081_roi} shows our ROI conditions, where our target is close to the Galactic plane, located in a region with many point-like and extended sources. We then fit our ROI model by maximizing the likelihood with the selected data. To do this, we free the normalization parameter ($\phi_0$) of all sources in the base model located within a radius of $4^\circ$ from our target, as well as the spectral index ($\Gamma$) of sources closer than $2^\circ$. Figure \ref{fig:HH8081_roi} also highlights 4FGL J1818.5-2036, an unassociated point-like source with $4\sigma$ detection significance in the 4FGL-DR4 catalog, separated by $\sim 0.2^\circ$ from IRAS 18162-2048. Considering the large 68\% containment angle of \textit{Fermi}-LAT ($\sim 5^\circ$ at $\rm 100 \, MeV$), which reflects the PSF, this source may be potentially associated with our target. We thus performed a fitting following the same methodology as before, but replacing 4FGL J1818.5-2036 by a point-like source at the exact coordinates of IRAS 18162-2048 \citep[see][]{IRAS_position}. This modification significantly increased the likelihood of the whole model, returning a logarithmic likelihood-ratio test of $\sim 42$. Therefore, from now on, 4FGL J1818.5-2036 is removed from our base source model and substituted by a point-like source centered in our MYSO coordinates, accounting for the high-energy emission emerging from the proximities of the protostellar jet (details of the new source can be found in Table \ref{tab:source_model}). The updated ROI fitted model will serve as the basis for all the following analysis.

%Due to the proximity between both sources, we substitute the J1818.5-2036 point-like source of the 4FGL catalog with another point source at the exact coordinates of IRAS 18162-2048 (see Table \ref{tab:coordinates}). We then fit our ROI model by maximizing the likelihood with the selected data. To do this, we free the normalization parameter of all sources in the base model located within a radius of $4^\circ$ from our target, as well as the spectral index of sources closer than $2^\circ$. Figure \ref{fig:HH8081_roi} shows our ROI conditions, where our target is extremely close to the Galactic plane, located in a region with many point-like sources and extended sources.

\subsection{Detection and TS maps}
\label{sect:detection}

 The PSF of the LAT instrument varies from $\sim 5 ^\circ$ at 100 MeV to $\sim 0.1^\circ$ at 100 GeV, limiting the spacial resolution of the instrument. \cite{Zucker_2020} established the distance to the source at $\rm (1400 \pm 70)\, pc$, projecting an angular extension of $\sim 0.3 ^\circ$ for the entire system. Despite the enormous size of this particular jet of $\rm{ \sim 10 \, pc}$ \citep{Masque_2015}, the angular size of our source is similar to the angular resolution at the highest energies, where the number of photo-events tends to be less abundant in a typically decreasing spectrum. Therefore, we expect the entire HH 80-81 system (hereafter HH 80-81) to be contained within a small angular region centered on IRAS 18162-2048, behaving as a point-like source. 
 
 Given the lack of any bright $\gamma$-ray source in the field of view, our ROI is dominated by the Galactic diffuse emission due to the proximity of HH 80-81 to the Galactic plane, which accounts for $\sim 80 \%$ of the detected photo-events in our analysis. Thus, the accuracy in the fitted model of this background emission is crucial for analyzing our object.
 At low energies, the containment area where we detect $\gamma$-ray photons provided by the target source is quite large, resulting in a considerable contamination from the background inside this region. Based on this, we studied the expected counts from the Galactic diffuse emission within the region covered by the PSF located in the same position as HH 80-81. Then, we find that a $\sim 1\%$ of variation in the diffuse emission corresponds to the same level of emission measured from the new source at $\rm 300 \, GeV$.
 %Our findings indicate that a variation of $\sim 1\%$ in the intensity of the Galactic diffuse emission at $\rm 300 \, MeV$ corresponds to emission detected from HH 80-81 at the same energies. 
 Therefore, we decided to ignore all data below $\rm 300\, MeV$ since we considered them highly affected and unreliable because of their sensitivity to the Galactic diffuse emission.
 % Final fit model
 As a result, following the methodology described in the beginning of Sect. \ref{sect:analysis}, we obtain a ROI fitted model that provides an initial approach to the source of interest (see Table \ref{tab:source_model}).

\begin{table}[]
\caption{Initial parameters for the source of interest obtained from the fitted ROI model.}
\centering
\begin{tabular}{@{}c@{}}
\toprule
\textbf{Source of interest}                                                       \\ \midrule
TS $_{E>300 \, {\rm MeV} } = 29.44$ \ | \ TS $_{E>100 \, {\rm MeV} } = 41.92$                \\
$N_{{\rm pred} , \ E>300 \, {\rm MeV}}  = 1445$ \ | \  $N_{{\rm pred} , \ E>100 \, {\rm MeV}}  = 4673$                      \\ \midrule
Power-law spectral model (see Eq. (\ref{eq:powerlaw}))                            \\ \midrule
$\Gamma = 2.62 \pm 0.12$                                                          \\
$\phi_0 =\left( 7.6 \pm 2.2 \right) \times 10^{-13} \, \rm MeV^{-1} cm^{-2} s^{-1} $ \\ \midrule
Point-like spatial model                                                          \\ \midrule
${\rm RA_{J2000}} = 274^\circ .80$                                                     \\
${\rm Dec_{J2000}} = -20^\circ .79$                                             \\ \bottomrule
\end{tabular}
\label{tab:source_model}
\end{table}

Once we have built a reliable dataset, we obtain several test statistic (TS) maps in order to display our source detection (Fig. \ref{fig:ts_maps}). We use the \textit{tsmap} method from \texttt{fermipy}, which computes the TS value for each spatial bin by adding a test source at each spatial position in the ROI calculating its amplitude and TS versus the location, in order to get a significance map. The remaining parameters are manually fixed. In our case, we have used the simplest test source possible: a point-like source whose spectrum is modeled with a power law described by Eq. (\ref{eq:powerlaw}),
\begin{equation}
    \phi (E) =  \phi_0 \left(  \frac{E}{E_0}  \right) ^{- \Gamma} \ {\rm MeV^{-1} \, cm^{-2} \, s^{-1} },
    \label{eq:powerlaw}
\end{equation}
where $\phi (E)$ describes the differential flux at a certain energy and $\phi_0$ is the event rate at the reference energy $E_0$. In this work, we fixed the reference energy to $ E_0 = \rm 1 \, GeV$ for our spectral models of HH 80-81. In addition, to compute the TS value in each spatial bin, we create a test source with $\Gamma_{\rm test} = 2.6$ based on the results of Table \ref{tab:source_model} from the ROI fitted model.

Figure \ref{fig:ts_maps} is created following this methodology. Panels (a) to (c) represent the significance of the counts excess at different minimum energy thresholds. These are computed by excluding our target source from the final fitted model. The source detection is achieved if the TS is over 25, which is equivalent to a $ 5 \sigma$ detection significance. Figure \ref{fig:ts_maps} shows that the source of interest is still detected when considering a lower energy limit of $\rm 500 \, MeV$. However, the emission above 700 MeV is not enough to reach the $5 \sigma $ detection threshold. On the other hand, panel (d) shows a flat TS map (this time, considering the presence of HH 80-81 in the ROI model) for a test point source with $\Gamma_{\rm test} = 2$, confirming the good quality of the fitted model.

\cite{Yan2022} initially detected a $\gamma$-ray excess in the same region between $\rm 100 \, MeV$ and $\rm 300 \, MeV$. Now, by using five more years of data and a more precise analysis, we have identified some incompatibilities. Analysing the dataset included in this work, we found that the count excess yields a TS of $\sim 29$ above $\rm 300 \, MeV$ ($\sim 42$ above $\rm 100 \, MeV$), reaching the $5 \sigma$ threshold to claim the source detection. This is notably lower than the $10 \sigma$ detection claimed by \cite{Yan2022} above $\rm 100 \, MeV$. Furthermore, the source is clearly detected above $\rm 500 \, MeV$, indicating a harder spectrum than that reported in \cite{Yan2022} ($\Gamma = 3.53 \pm 0.11$). These discrepancies can be attributed to the energy range used in the initial detection. As we have calculated, the Galactic diffuse emission is dominating the lowest energies, where the first detection was achieved with significantly less exposure time, suggesting that the original detection was heavily affected by the residuals of the Galactic template.

\subsection{Morphological analysis}
\label{sect:morphology}

%\begin{figure}
%    \centering
%    \includegraphics[width=82mm]{extension.pdf}
%    \caption{Analysis of the extension of HH 80-81. \textbf{Top:} White contours represent part of the extension models that we used to fit HH 80-81 source. Green shows the best extension model that maximize the likelihood compared with the PSF for 300 MeV (yellow). \textbf{Bottom:} Relation between likelihood  and radius of the extension model. \textcolor{red}{Although the likelihood of the best extension model is higher than the likelihood of a point-like source, the difference is less than $3\sigma$, not enough to be significant. (Explain how TS\_ext is calculated may be better!)}}
%    \label{fig:extension}
%\end{figure}

The low spatial resolution of \textit{Fermi}-LAT restricts morphological studies on small sources like HH 80-81. However, accelerated particles can reach large regions with high ISM densities, producing $\gamma$-ray excess along enormous areas that can be detected as extended sources by \textit{Fermi}-LAT. To constrain the extension of the source of interest, we employ the \textit{extension} analysis algorithm in the \texttt{fermipy} package.

We build 21 radial disk models, varying the radius from $0.01 ^\circ$ to $1.0^\circ$, all centered on the same coordinates of IRAS 18162-2048 (Table \ref{tab:coordinates}). For each uniform disk  model, we fit the spectral parameters of the sources by maximizing the likelihood. Then, the global likelihoods of all the models are compared to determine the significance of the extended model as a function of the source radius. This significance is calculated as the square root of the $\Delta {\rm TS}_{\rm ext}$, obtained through Eq. (\ref{eq:extension}), where we compare the global likelihood of each extension model ($\mathcal{L}_{\rm extended}$) with the likelihood obtained for a point-like source ($\mathcal{L}_{\rm point-like}$).
\begin{equation}
    \Delta {\rm TS} _{\rm ext} = -2 \, \left( \ln \mathcal{L}_{\rm extended} - \ln \mathcal{L}_{\rm point-like} \right).
    \label{eq:extension}
\end{equation}

The extended model that best fits to the detected $\gamma$-ray emission in the region has a radius of $(0.25 \pm 0.05) ^\circ$, resulting in a hint of $\Delta{\rm TS_{ext}} \approx 10$, which does not reach the commonly used $5\sigma$ threshold for claiming an extended source. Since our source radius is significantly lower than our PSF at low energies, where the majority of photo-events are expected to be, we can assume that our source of interest behaves as a point-like source. Consequently, we continue our analysis without considering any extension.

\subsection{Spectral analysis}
\label{sect:sed}

Based on the \textit{sed} method of \texttt{fermipy}, we calculate the high-energy spectral energy distribution (SED) of our source of interest. Here, the dataset is divided into several energy bins where the normalization parameter in Eq. (\ref{eq:powerlaw}) is fitted to maximize the likelihood of the model in each bin, keeping the rest of parameters fixed. In this case, we use 4 energy bins per decade, ranging from 300 MeV to 100 GeV, and we fix the spectral index of HH 80-81 to 2.6, assuming that the overall spectral slope that we obtain in Table \ref{tab:source_model} is stable and close to the real value. We also free the normalization parameters regulating the Galactic diffuse emission and the isotropic emission, as both dominate our ROI. Additionally, the selected shape of the spectrum, starting from a power-law, may play a remarkable roll in performing the SED analysis. Depending on the shape of the spectrum, we are forcing the analysis software to fit the emission of the putative source. Thus, we have to determine which model fits better to the detected emission.

%Hitherto
Heretofore, the detected emission has been initially modeled as a power law based on simplicity arguments. However, we can use other spectral shapes to describe a point like source, such as the log-parabola spectrum in Eq. (\ref{eq:logparabola}) 
\begin{equation}
    \phi (E) = \phi_0 \left(  \frac{E}{E_0} \right) ^{- \Gamma - \beta \ln \left( \nicefrac{E}{E_0}  \right)} \ {\rm MeV^{-1} \, cm^{-2} \, s^{-1} },
    \label{eq:logparabola}
\end{equation}
or a power law with an exponential cutoff model (hereafter exponential cutoff) similar to that in Eq. (\ref{eq:cutoff}), where additional parameters affect to the curvature of the spectra. The power-law expression is nested in these two spectral shapes, allowing the comparison between all these different models. Employing the \textit{curvature} algorithm of \texttt{fermipy}, we construct the TS significance for these three models by comparing the maximum likelihood achievable for each model in Eq. (\ref{eq:curvtest}).
\begin{equation}
    \Delta{\rm TS_{model}} = -2 \left(  \ln \mathcal{L}_{\rm PowerLaw} - \ln \mathcal{L}_{\rm model}  \right).
    \label{eq:curvtest}
\end{equation}
In Eq. (\ref{eq:curvtest}), the suffix `\textit{model}' refers to the log-parabola or exponential cutoff models compared to the power-law shape. Table \ref{tab:curvtest} shows the comparison between these three models. Based on the results, the low $\rm \Delta TS$ values describing the goodness of the different fits suggest that the high-energy spectrum of the source can be accurately described by a power-law function.

\begin{table}
    \centering
    \caption{Comparison of the different models for computing the spectrum of the $\gamma$-ray detection.}
    \label{tab:curvtest}
    \begin{tabular}{@{}lcc@{}}
        \toprule
        \multicolumn{1}{c}{\textbf{Spectral shape}} & \textbf{LogLikelihood} & $\rm \mathbf{ \Delta TS _{model}}$ \\ \midrule
        Power-law                                   & 9417073.0             & --                      \\
        Log Parabola                                & 9417073.8             & 1.7                     \\
        Exponential Cutoff                          & 9417073.6             & 1.3                     \\ \bottomrule
    \end{tabular}
\end{table}

\begin{figure}
    \centering
    \includegraphics[width=82mm]{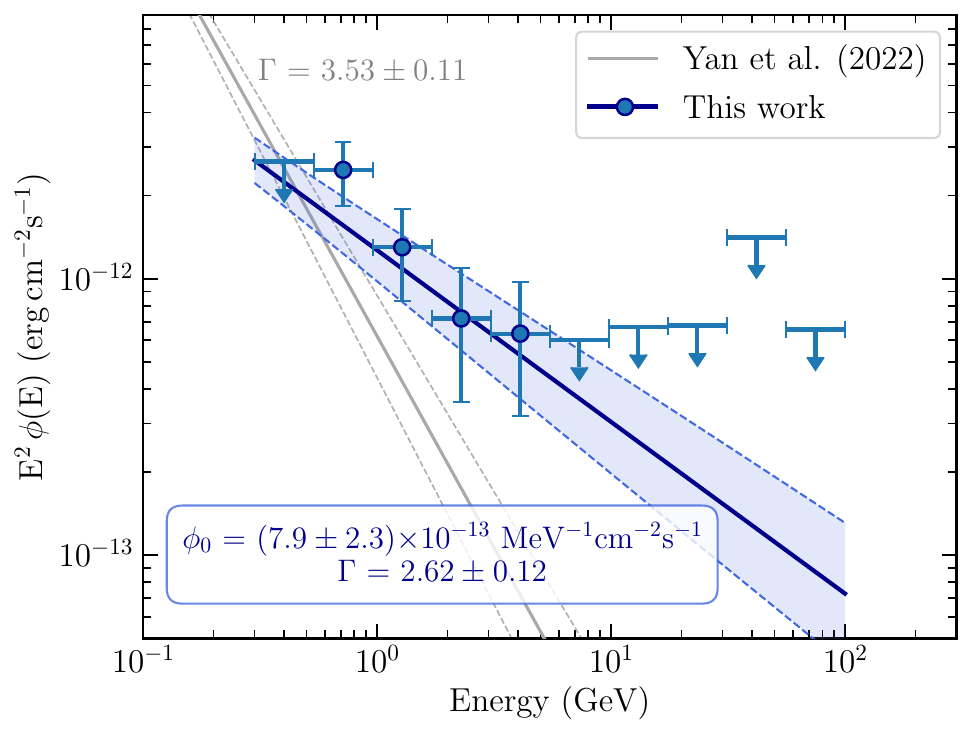}
    \caption{SED of the source of interest. Blue circles represent the significant points obtained from our analysis, while blue arrows show the upper limits for those energy bins that do not reach a minimum TS value of 4. The blue line shows the power-law fit resulting from this work and the blue box shows the parameters described in Eq. (\ref{eq:powerlaw}). The previous analysis performed by \cite{Yan2022} is shown in grey.}
    \label{fig:sed_fermi}
\end{figure}

Figure \ref{fig:sed_fermi} shows the spectrum obtained for our detection. Comparing with previous results, Fig. \ref{fig:sed_fermi} shows a clear discrepancy in the slope of both spectra. As mentioned before, we attribute these differences to the Galactic diffuse emission, as \cite{Yan2022} has detected the source only in the low energy regime (bellow $\rm 300 \, MeV$), where the Galactic diffuse background is heavily dominant over our source. These new results, with 15 years of collected data, allow the detection of our source of interest up to higher energies, describing a harder power-law spectrum. Every point plotted in the Fig. \ref{fig:sed_fermi} has more than 10 predicted events, and upper limits are calculated with a $95\%$ confidence level for all the bins that do not reach the minimum of predicted events nor the critical TS value of 4.

\subsection{Variability analysis}
\label{sect:variabilty}

\begin{figure}
    \centering
    \includegraphics[width=78mm]{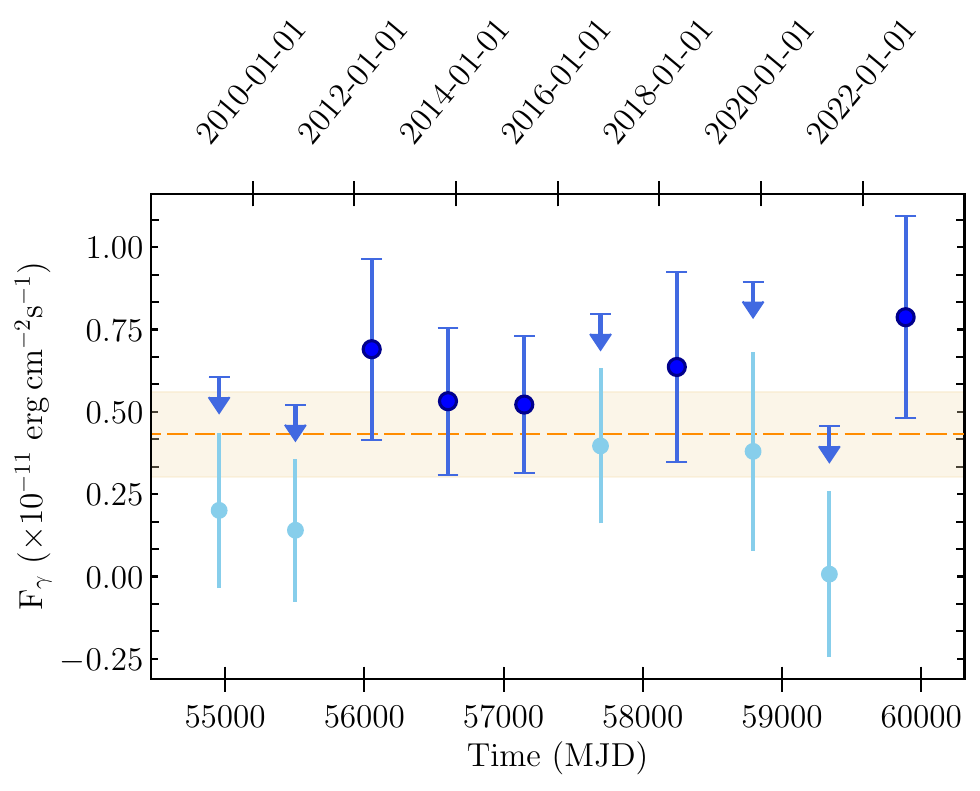}
    \caption{Variability of the $\gamma$-ray energy flux of the source of interest over 15 years of observations. Dark blue points indicate the $\gamma$-ray flux during periods where the source detection exceeds the $2\sigma$ threshold. For non-detected time intervals, upper limits are shown with arrows, representing the $95\%$ confidence level. Besides, light blue points show the flux derived using those non significant points, providing an indication of the behavior of the source during those periods. The horizontal dashed line and shaded area represent the detected energy flux over the 15 years of observations and the $1\sigma$ uncertainty area, respectively.}
    \label{fig:lightcurve}
\end{figure}

The recent bursts detected using radio, submilimiter, and IR wavelengths in the cases of NGC6334I-MM1 \citep{Hunter_17}, S255 NIRS3 \citep{Caratti_o_Garatti_17}, G358 \citep{Stecklum_2021}, M17 MIR \citep{Chen_2021}, G11.497-1.485 \citep{Bayandina_2024}, and G323 \citep{Wolf_2024} demonstrate that MYSOs, as well as protostars that are still accreting mass, can gain vast masses in short periods of time due to disk instabilities, increasing the jet activity and emitting new knots \citep[e.g.][]{Marti_95, Cesarone_18, Fedriani_2023}. These powerful events lead to flare episodes that can be detected in $\gamma$ rays to definitely confirm the association between a \textit{Fermi}-LAT source and a protostellar jet as a counterpart \citep[see][]{deOna_2023}. Unfortunately, since the beginning of \textit{Fermi}-LAT's operations, no flares have been reported from HH 80-81. 

Hence, we do not expect to find any variability in the intensity of the $\gamma$-ray emission collected so far. The \textit{lightcurve} method applied in this work involves splitting the initial dataset into 10 equivalent time bins. With all the spectral parameters of all the sources in our ROI fixed, we fit the normalization parameters of the $\gamma$-ray source, the Galactic diffuse emission, and the isotropic emission in order to maximize the likelihood. Finally, we save the TS detection value and the energy flux emitted in each time bin. 

Figure \ref{fig:lightcurve}, produced following this methodology, demonstrates the flux stability of our detection. The figure shows the integrated spectrum over the energy range used in the analysis, from 300 MeV to 100 GeV, resulting in the integral $\gamma$-ray flux ($F_\gamma$). Due to the marginal detection of the source in short time periods, we obtain important uncertainties for each time bin and the low-flux time periods do not reach the detection threshold. However, deviations from the average energy flux are smaller than $2\sigma$, consistent with our expectations of no variability in this source.

\section{Source Identification}
\label{sect:association}

The position of HH 80-81 appears to be slightly shifted from the peak position in the TS maps shown in Fig. \ref{fig:ts_maps}. In addition, the spatial resolution of \textit{Fermi}-LAT makes it challenging to definitely associate $\gamma$-ray sources with unique counterparts, as the large containment area of the instrument may content a vast population of galactic and extragalactic objects. However, since HH 80-81 has not been reported flaring in other wavelengths, we can only try to associate the $\gamma$-ray detection presented in Sect. \ref{sect:analysis} via positional arguments.

\begin{table}
\centering
\caption{Comparison between the coordinates of IRAS 18162-2048 \citep{IRAS_position} obtained with the Atacama Large Milimiter Array (ALMA) and the best position obtained for the \textit{Fermi}-LAT source.}
\label{tab:coordinates}
\begin{tabular}{@{}cccc@{}}
\toprule
          & \multicolumn{1}{c}{\textbf{RA}$\rm _{\mathbf{J2000}}$} & \multicolumn{1}{c}{\textbf{Dec}$\rm _{\mathbf{J2000}}$}   & \multicolumn{1}{c}{\textbf{Separation}} \\ \midrule
ALMA      & $\rm 18^h 19^m 12^s.1$  & $\rm -20^\circ 47' 31''$ & --                             \\
\textit{Fermi}-LAT & $\rm 18^h 19^m 30^s \pm 20^s$ & $\rm -20^\circ 51' \pm 5'$  & $\rm 0.11^\circ$          \\ \bottomrule    
\end{tabular}
\end{table}

\begin{figure}
    \centering
    \includegraphics[width=80mm]{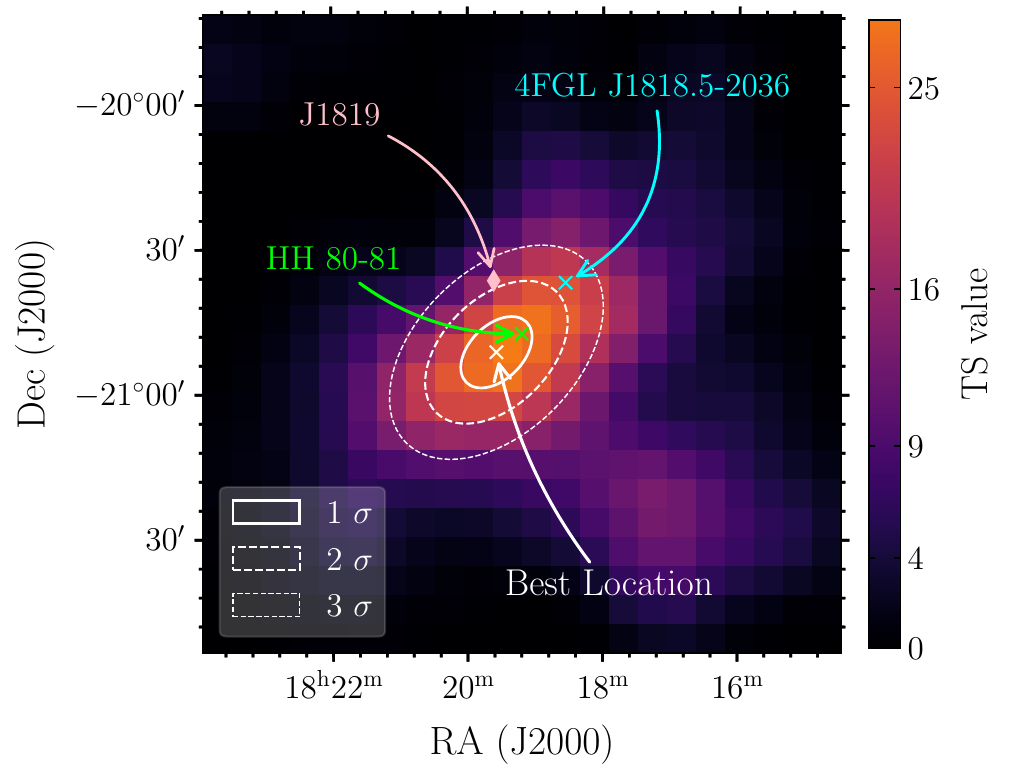}
    \caption{Best location of the $\gamma$-ray detection that maximize the likelihood of our model. White contours represent the uncertainty of the modeled position, while the color map shows the TS of the excess related to HH 80-81. The cyan mark indicates the position of 4FGL J1818.5-2036, which was removed from the initial model due to its relation to the HH 80-81 system.}
    \label{fig:location}
\end{figure}

First of all, we applied the \textit{localize} method of \texttt{fermipy} to fit the best location for the source of interest that we have analyzed. This method fits the best position of the putative source, returning the best position of a point-like source for explaining the $\gamma$-ray excess as well as the uncertainties in the position. Table \ref{tab:coordinates} shows the final value after this process while Fig. \ref{fig:location} illustrates the source location. Since IRAS 18162-2048 is located within the $1\sigma$ area, the position of our $\gamma$-ray source is compatible with HH 80-81. Additionally, the initial 4FGL source J1818.5-2036 stays within the $3\sigma$ region.

In order to clarify the origin of the detected $\gamma$-ray emission, we investigated all the potential counterparts populating the $3\sigma$ region. We checked the ATNF Pulsar Catalogue \citep{Pulsar_cat}, the SNRcat \citep{SNRcat}, the Green's Galactic supernova remnants catalogue \citep{SNR_Green}, and several active Galactic nuclei (AGNs) catalogues compiled by \cite{Kauffmann_AGNcat} and \cite{Assef_AGNcat}, confirming that there are no other well-characterized candidates spatially coincident with the $3\sigma$ location of the \textit{Fermi} emission. Additionally, we reviewed the Radio Fundamental Catalog\footnote{\href{https://astrogeo.org/sol/rfc/rfc_2024b/}{https://astrogeo.org/sol/rfc/rfc\_2024b/}}, taking into account all the compact bright radio sources in the region that can be affected by the high extinction of the Galactic plane complicating their classification for these catalogs. We just identified ICRF3 J1819-2036 ($\rm RA_{J2000}=18^h19^m36^s.9 ,\ Dec_{J2000} = -20^\circ 36'32''$) within the $\rm 3 \, \sigma$ area (see Fig. \ref{fig:location}).

ICRF3 J1819-2036 (hereafter J1819) is a bright radio source listed in the ICRF3 catalog as a reliable VLBI calibrator \citep{ICRF3_paper}, which has been used as a phase calibrator in some papers of the literature \citep[e.g.][]{J1819_2020, Immer_22}. It is a rapid variable radio-source, presenting significant flux variations in few hours. Nevertheless, no flares have been reported despite it has been observed since 1990 \citep{J1819_1994}. Very Large Array (VLA) has observed J1819 four times from 2015 to 2018 in the S and X band ($\rm 2.3\, GHz$ and $\rm 8.7 \,GHz$, respectively) showing a point-like morphology that can be compatible with an AGN \citep{Petrov_19}. The photometric radio spectrum is notably bright (of the order of $\rm \sim 0.1\, Jy$ in both bands), with a negative spectral index indicative of non-thermal synchrotron emission. Furthermore, the null proper motion detected in those three years of observation is consistent with an extragalactic source. 

Since quasars and blazars can be $\gamma$-ray emitters, we characterized as much as possible J1819 in order to determine whether its presence, close to HH 80-81, could explain our detection. The source is faintly visible in the far-IR band and undetectable at near-IR and optical wavelengths, likely due to the Galactic extinction. We examined the soft and hard X-ray emission observed by XMM-Newton in the region \citep{Pravdo_2004}. In X-rays, the expected extinction is much lower than optical or IR absorption, just $\rm \sim 2 \, mag$ in the soft band and $\rm \sim 0.1 \, mag$ in hard X rays \citep{X-ray_absroption}. Therefore, if J1819 is emitting hard X-rays, it should be detectable. However, there are no detection of this source, constraining the hard X-ray emission to be $\rm <10^{-14} \, erg \, s^{-1} cm^{-2}$, which may be incompatible with the overall spectrum for IC scattering under certain conditions. We also checked the HESS Galactic Plane Survey \citep{HESS_Galactic_survey}, that observed at J1819's position without any hint of detection. 
Lastly, the stable variability curve detected with \textit{Fermi}-LAT (see Sect. \ref{sect:variabilty}) suggests that the $\gamma$-ray source might not be an AGN, as most of such objects detected by \textit{Fermi} present significant flux variations \citep{Variability_blazars_2020}.

Also, \cite{Pravdo_2004} report the presence of MRR 12 ($\rm RA_{J2000}=18^h19^m10^s.4 ,\ Dec_{J2000} = -20^\circ 46'57''$) and MRR 32 ($\rm RA_{J2000}=18^h19^m21^s.8 ,\ Dec_{J2000} = -20^\circ 45'35''$), two interesting X-ray sources located close to the jet. Both are associated with compact nebular objects (typically star-forming objects) in IR wavelengths, and included in the GGD catalog \citep{GGD_catalog} of HH objects. MRR 12 also appears as a possible non-thermal emitter, given its negative spectral index in radio frequencies. However, both objects are classified as Class III pre-main sequence stars \citep[see][]{Pravdo_2009}, in the weak T-Tauri phases. Therefore, there is no reason to consider them as potential $\gamma$-ray emitters.

Considering all of the above, HH 80-81 protostellar jet stands out as the main candidate to explain the $\gamma$-ray detection.
We examined all the Simbad sources surrounding IRAS 18162-2048 within a circular region of $\sim 0.44 ^\circ$, which is the semimajor axis of the $3\sigma$ area. Most of the radio sources and X-ray sources identified in the region are related to HH 80-81 or to young stars, not presenting any characteristic that could indicate non-thermal emission.
IRAS 18162-2048 is located in the $\rm 1 \sigma $ region while J1819 is located outside the $\rm 2 \sigma$ region (see Fig. \ref{fig:location}), even when selecting only \texttt{FRONT} events. Moreover, both radio spectrum \citep{Marti_1993, Rodriguez-Kamenetzky_2017} and polarized emission \citep{Carrasco-Gonzalez_2010} provide strong evidence for non-thermal emission in HH 80-81. It is also well-detected in soft and hard X-rays, showing a hard power-law emission \citep{Rodriguez-Kamenetzky_2019} compatible with our $\gamma$-ray emission.

Finally, \cite{Munar_11} determined the reliability of associations between $\gamma$-ray sources and MYSOs, concluding that $\sim 70\% $ of the associations between \textit{Fermi}-LAT sources and MYSO counterparts are authentic.
Therefore, we should expect some $\gamma$-ray emission from the most clear protostellar jet with non-thermal emission. All these reasons lead us to consider HH 80-81 as the main candidate to be powering the detected high-energy emission analyzed in Sect. \ref{sect:analysis}.

%\textcolor{red}{ Esto no sé si quitarlo.} Finally, we examined all the Simbad sources surrounding IRAS 18162-2048 within a circular region of $\sim 0.44 ^\circ$, which is the semimajor axis of the $3\sigma$ area. Most of the radio sources and X-ray sources identified in the region are related to the protostellar jet or young stars, not presenting any characteristic that could indicate non-thermal emission.

\section{Origin of the $\gamma$-ray emission}

\label{sect:gammaprod}

%Based on the results of the Sect. \ref{sect:analysis}, where we performed a detailed analysis of the high energy emission of HH 80-81, we can infer the particle distribution responsible for the $\gamma$-ray emission. We use \texttt{naima}\footnote{\href{https://naima.readthedocs.io/en/latest/radiative.html}{https://naima.readthedocs.io/en/latest/radiative.html}} \citep{naima} to fit a radiative model to our data based on a determined particle distribution. Therefore, we must estimate which radiative processes dominate in our source. 

\begin{figure}
    \centering
    \includegraphics[width=80mm]{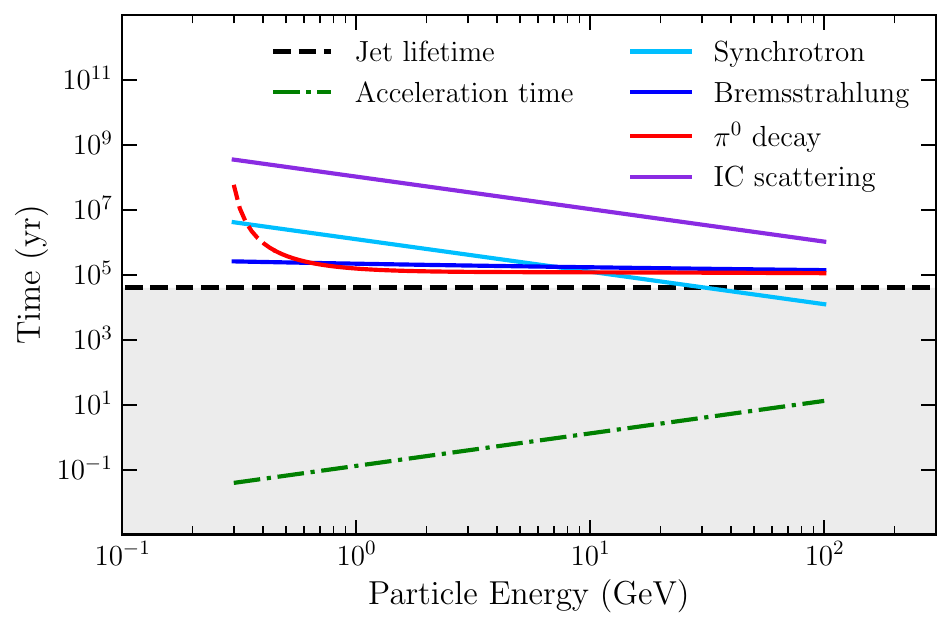}
    \caption{Relevant timescales for the HH 80-81 system. Solid lines show the cooling timescales for different emission models. The black dashed line represents the estimated age of the jet, and the dash-dotted green line indicates the minimum time required to accelerate particles up to different energies based on the jet properties.}
    \label{fig:cooling-times}
\end{figure}

To understand the radiative spectrum from Sect. \ref{sect:sed} flowing out from the protostellar system, we determine the most effective radiation mechanisms considering different populations of particles under the ambient and internal conditions of HH 80-81. Figure \ref{fig:cooling-times} shows cooling timescales for the usual radiative mechanisms for relativistic protons and electrons estimated with Eqs. (10) -- (14) from \cite{Bosch-Ramon_2010} and \texttt{Gamera}\footnote{\href{https://github.com/libgamera/GAMERA}{https://github.com/libgamera/GAMERA}} \citep{Gamera_15, Gamera_2022}. 
The IC scattering is calculated using an IR density of $\rm 3 \, eV \, cm^{-3}$, consistent with \cite{Bosch-Ramon_2010} and derived from the IR luminosity of IRAS 18162-2048\footnote{\href{https://tools.ssdc.asi.it/SED/sed.jsp?\&ra=274.79917\&dec=-20.79306\&q=IRAS\%2018162-2048}{https://tools.ssdc.asi.it/SED/sed.jsp?\&ra=274.79917\&dec=-20.7930 6\&q=IRAS\%2018162-2048}} contained in a spherical region of $\rm 2.5 \, pc$ where HH 80, HH 80N, and HH 81 are located. The synchrotron cooling time is calculated for an electron population within a magnetic field of $\rm 0.1 \, mG$ \citep{Carrasco-Gonzalez_2010, Rodriguez-Kamenetzky_2019}. The relativistic Bremsstrahlung radiative process and the $\gamma$-ray emission from inelastic proton collisions ($\pi^0$ decay) are calculated using a density value of 100 $\rm cm^{-3}$ for the ambient medium \citep{Bally_23}. We adopted a  jet lifetime of $\rm 4 \times 10^4 \, yr$, obtained in \cite{Qiu_2019}, and the timescale for particle acceleration is computed assuming a Bohm diffusion coefficient and a shock velocity of $\rm 400 \, km \, s^{-1}$ \citep{Marti_95, Marti_98, Masque_2015, Bally_23}.
As Fig. \ref{fig:cooling-times} shows, the most effective mechanisms for loosing energy are synchrotron and relativistic Bremsstrahlung emission, and proton-proton collisions for a leptonic or a hadronic model, respectively. Regarding the $\gamma$-ray band, we consider relativistic Bremsstrahlung and $\pi^0$ decay as the most possible radiative mechanisms, taking into account that IC scattering is negligible due to the low photon density.

%The key objective of this type of analysis is to distinguish between leptons and hadrons to empirically confirm effective CR production within a region. 

The LAT data from Fig. \ref{fig:sed_fermi} are fitted using the \texttt{naima}\footnote{\href{https://naima.readthedocs.io/en/latest/radiative.html}{https://naima.readthedocs.io/en/latest/radiative.html}} tool \citep{naima} considering the two dominant mechanisms: a leptonic population producing $\gamma$ rays via relativistic Bremsstrahlung; and a hadronic population, whose emission is due to proton-proton inelastic collisions.
The properties of the medium where particles are accelerated are also important in order to determine the number of particles needed to produce the $\gamma$-ray emission, no matter if the origin is hadronic or leptonic. Several studies address the density of the outflow gas \citep{Qiu_2019} and provide results on the specific density of the HH 80 and HH 81 knots based on X-ray emission \citep[see][]{Pravdo_2004, Rodriguez-Kamenetzky_2017}. \cite{Heathcote_98} and \cite{Masque_2015} estimate the density of the jet of the order of $\rm \sim 1000 \, cm^{-3}$, while \cite{Bally_23} provide $\rm 100 \, cm^{-3}$ for the ISM density given the position of HH 80-81 in the outskirts of L291. 
Given that the two high-energy radiative mechanisms under consideration directly depends on the density of the ambient medium \citep[see][]{Bremsstrahlung_dep, Pion_dep}, the integrated energy of the particle distribution is inversely proportional to ambient density. Accordingly, we adopt a density value of $ 100 \, \rm cm^{-3}$, the most restrictive scenario, which demands higher particle energy to explain the detected $\gamma$-ray emission.

%The expected particle SED originated by DSA in the ultra-relativistic regime is a power-law spectrum with a spectral index of $\sim 2$ for a monoatomic gas. This represents the hardest spectrum achievable via DSA, with softer spectra expected at lower energies. Although protons are considerably heavier than electrons and do not fall strictly within the ultra-relativistic regime, minor changes in the spectral index do not significantly affect the integrated particle energy. Thus, we compute the particle distribution of protons as a power law with an exponential cutoff, fixing $\Gamma =2$. We allow only the normalization parameter and the cutoff energy to vary, as shown in Eq. (\ref{eq:cutoff}). In contrast, electrons are more sensitive to external conditions, and their original particle distribution is easily modulated. Therefore, we do not assume any spectral index. Instead, we model their distribution with a power law freeing $\Gamma$ and $\phi_0$ for simplicity.

To model both hadronic and leptonic scenarios, we employ a particle distribution characterized by a power law with an exponential cutoff, as described in Eq. (\ref{eq:cutoff}), freeing the normalization parameter ($\phi_0$).

\begin{equation}
    \phi (E) = \phi_0 \left( \frac{E}{\rm 1 \, GeV} \right) ^{-\Gamma} \exp \left( - \frac{E}{E_{\rm cutoff}} \right) \ {\rm MeV^{-1} \, cm^{-2} \, s^{-1} }.
    \label{eq:cutoff}
\end{equation}

Figure \ref{fig:cooling-times} shows an almost null dependence of the relativistic Bremsstrahlung and $\pi^0$ decay processes on particle energy. Therefore, we adopt the same spectral index for the particle distribution as that of the photo-spectrum ($\Gamma_{\rm particle} = 2.62$). Furthermore, from Fig. \ref{fig:cooling-times} we observe that if cutoff energy ($E_{\rm cutoff}$) is determined by radiative cooling, the maximum energy reached by electrons will be lower than the cutoff energy of protons, as a consequence of the lower cooling time for synchrotron radiation at the highest energies.
Using Eq. (17) from \cite{Bosch-Ramon_2010}, we estimate a maximum particle energy of $E_{\rm cutoff, \, e} \approx 3 \, {\rm TeV}$, resulting from synchrotron cooling under the specific conditions of HH 80-81. As well, applying Eq. (20) from \cite{Bosch-Ramon_2010}, we obtain the maximum particle energy achievable via $\pi^0$ decay cooling for a density of $\rm 100 \, cm^{-3}$, resulting in $E_{\rm cutoff, \, p} \approx 12 \, {\rm TeV}$.
Additionally, \cite{Araudo_2021} also predicts that protons reach higher energies than electrons when the cutoff energy is determined by the escape of high-energy protons from the upstream region of the shock. Since $E_{\rm cutoff, \, p} > E_{\rm cutoff, \, e}$ and $E_{\rm cutoff, \, e}$ is well above our region of interest, the particle distribution approximates a power-law within the studied energies.

%\textcolor{red}{For modeling the hadronic population, we use a proton particle distribution described by a power-law function with an exponential cutoff, fixing the spectral index $\Gamma _{\rm p} = 2$ (expected from DSA). We allow only the normalization parameter and the cutoff energy to vary, as shown in Eq. (\ref{eq:cutoff}). In contrast, electrons are easily affected by external effects like energy losses, modifying the spectrum of the parent particle distribution. Consequently, we do not assume any spectral index. Instead, we model their distribution with the power law from Eq. (\ref{eq:powerlaw}) freeing $\Gamma _{\rm e}$ and $\phi_0$ for simplicity.}

\begin{figure}
    \centering
    \includegraphics[width=80mm]{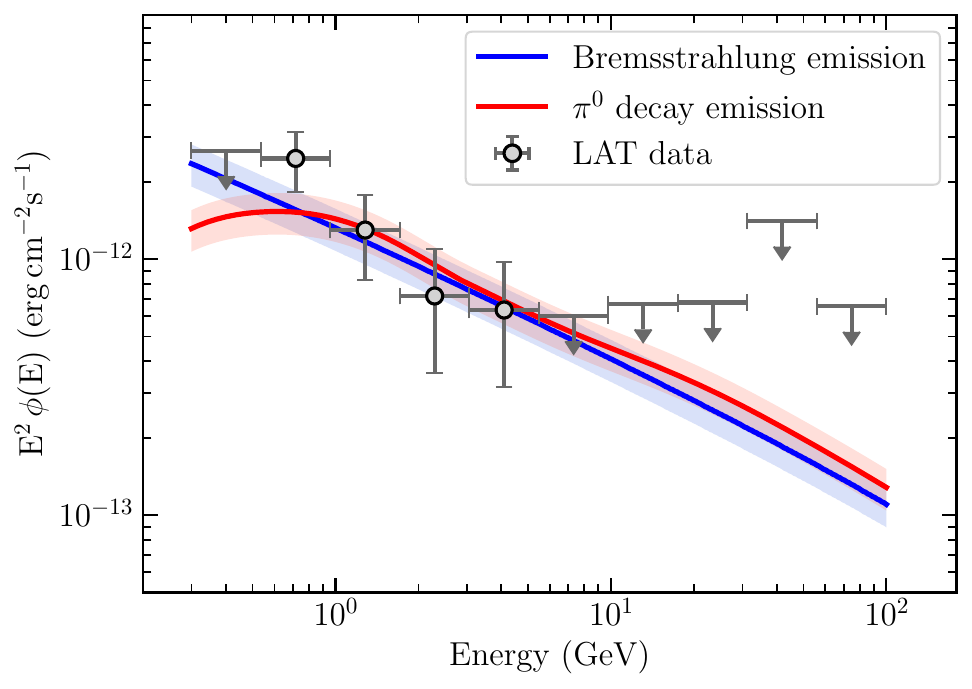}
    \caption{SED fitting for leptonic and hadronic models of $\gamma$-ray production. Grey points represent the \textit{Fermi}-LAT emission from Fig. \ref{fig:sed_fermi} obtained in this work. Shaded areas show the $1\sigma$ error for both fittings.}
    \label{fig:naima_sed}
\end{figure}

Figure \ref{fig:naima_sed} presents the radiative spectrum produced by both leptonic and hadronic fits. In principle, both models are able to explain the high energy emission detected by \textit{Fermi}-LAT. The assumed spectral index is softer than that predicted by DSA within ultrarelativistic regime ($\Gamma = 2$), but remains relatively close.

%\textcolor{red}{Figure \ref{fig:naima_sed} presents both leptonic and hadronic fits. In principle, both models are able to explain the high energy emission detected by \textit{Fermi}-LAT. The cut-off energy ($E_{\rm cutoff}$) where the proton spectrum sharply decreases is situated at $\rm \sim 10 \, GeV$, comparable to some values provide in Table 3 of \cite{Araudo_2021}. However this value might change significantly improving the detection in the most energetic part of the SED. Taking into account that the cooling time for Bremsstrahlung mechanism keeps almost constant with energy (Fig. \ref{fig:cooling-times}), the electron population show a spectral index of $\Gamma_{\rm e} = 2.72 \pm 0.20$, consistent with the spectral photo-index from Fig. \ref{fig:sed_fermi}. Additionally, we also expect a feature in the spectrum at $\rm \sim 30 \, GeV$, where the synchrotron cooling time begins to be shorter than the jet lifetime. At higher particle energies, the $\gamma$-ray production of a leptonic population will be suppressed since synchrotron radiation in radio wavelengths is more effective.}

If the protostellar jet is composed by a single electron population responsible for all our $\gamma$-ray emission, we can estimate the synchrotron emission assuming a magnetic field of $\rm 0.1 \, mG$. However, when comparing the synchrotron spectrum predicted by our leptonic distribution with ancillary data from \cite{Marti_1993} and \cite{Vig_2018}, we observe important discrepancies. Radio observations situate synchrotron radiation two orders of magnitude weaker than we expect. These results might be attributed to a synchrotron self-absorption in the inner regions of the jet or the contribution of accelerated protons to the $\gamma$-ray spectrum. Furthermore, variations in the ambient density and the magnetic field strength could also account for these differences. Therefore, the observed discrepancies may not be physically significant, and the considerable degeneracy intrinsic to the problem prevents us from constraining any specific characteristic of the medium.

\begin{table}
    \small
    \centering
    \caption{Integrated energy of the particle distribution and injection time required to accelerate all the particles of the distribution.}
    \label{tab:naima}
    \begin{tabular}{@{}lcc@{}}
    \toprule
    \multicolumn{1}{c}{\textbf{Model}}     & \begin{tabular}[c]{@{}c@{}}\textbf{Energy} $(\rm erg)$ \\  $\left( \times \frac{n}{100 \, \rm{cm^{-3}}} \right)^{-1}$\end{tabular}      & \begin{tabular}[c]{@{}c@{}}\textbf{Injection} \\  \text{\textbf{time}} $(\rm yr)$\end{tabular} \\ \midrule
    Bremsstrahlung                         & $\left( 1.4 \pm 0.3 \right) \times 10^{46}$                 & $\left( 8.9 \pm 1.7 \right) \times 10^2$                                                               \\
    Pion decay                             & $\left( 2.9 \pm 0.5 \right) \times 10^{47}$ & $\left( 1.9 \pm  0.4 \right) \times 10^4 $                                                               \\ \bottomrule
    \end{tabular}
\end{table}

%\begin{figure*}
%    \centering
%    \includegraphics[width=130mm]{mwlMap.pdf}
%    \caption{Comparison between $\gamma$-ray emission and molecular clouds in the region of HH 80-81. The color map in the background shows a high-resolution IR map from the band A of the MSX experiment \citep{MSX_experiment}, which is used to locate HH 80-81 (magenta square). The warm color map displays a detection map based on the square root of the TS (similar to Fig. \ref{fig:ts_maps}). Green contours indicate the column density of the molecular gas in the vicinity of HH 80-81. \textcolor{red}{I will update the figure to show the other Fermi-LAT sources of the complex and the global TS excess to show also the correlation in the northern part.}}
%    \label{fig:mwlMap}
%\end{figure*}

\begin{figure*}
    \centering
    \includegraphics[width=130mm]{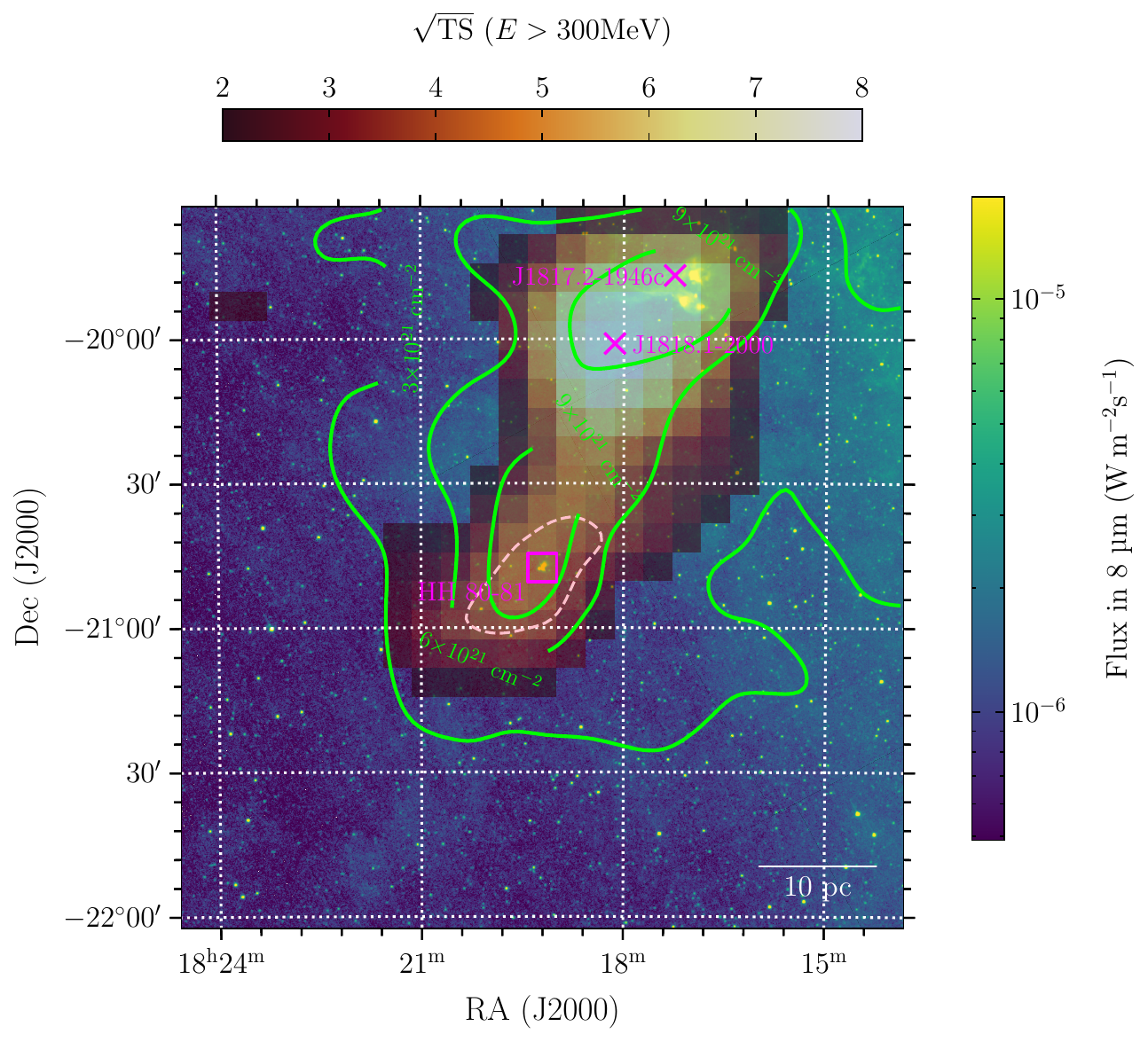}
    \caption{Comparison between $\gamma$-ray emission and molecular clouds in the region of HH 80-81. The color map in the background shows a high-resolution IR map from the band A ($\rm 8 \, \mu m$) of the MSX experiment \citep{MSX_experiment}, which is used to locate HH 80-81 (magenta square). Green solid contours indicate the column density of the molecular gas in the vicinity of HH 80-81. The warm color map displays a detection map based on the square root of the TS related to the three 4FGL sources populating the molecular complex, with magenta crosses locating the other two 4FGL sources apart from HH 80-81. The dashed contour indicate the $5\sigma$ detection area of HH 80-81 (see Fig. \ref{fig:ts_maps}).
    }
    \label{fig:mwlMap}
\end{figure*}

Integrating the particle distribution responsible for the emission of Fig. \ref{fig:naima_sed} from 300 MeV to 100 GeV provides the total energy accumulated by the accelerated particles of the source. Table \ref{tab:naima} shows the energy obtained for both models in function of the ISM density ($n$).  Based on a mass loss rate of $\rm \sim 10^{-5} \, M_\odot \, yr^{-1}$ \citep{Anez_2020} and the velocities of the outflow in the central regions of $\rm \sim 2000 \, km \, s^{-1}$ reported by \cite{Bally_23}, we estimate the jet kinetic luminosity to be of the order of $\rm \sim 10^{37} \, erg \, s^{-1}$. Assuming that 5\% of the kinetic energy of the jet is transformed into particle acceleration \citep{Araudo_2021} and using the Eq. (\ref{eq:injec_time}), we calculate the time required to accelerate the particles up to the necessary energy to produce the observed $\gamma$-ray emission (Table \ref{tab:naima}).
\begin{equation}
    t_{\rm injection} = \frac{E_{\rm total}}{\eta L_{\rm jet}},
    \label{eq:injec_time}
\end{equation}
where $\eta \approx 5\%$ is the acceleration efficiency, $E_{\rm total}$ is the integrated energy of the particle distribution (starting from $\rm 300 \, MeV$), and $L_{\rm jet}$ is the kinetic power of the jet.

In the lowest density regime ($100 \, \rm cm^{-3}$), the injection time for proton-proton collision is of the order of $\rm \sim 10^4 \, yr$, very similar to the estimated lifetime of the HH 80-81 protostellar jet \citep[$\sim 4 \times 10^4 \, \rm yr$;][]{Qiu_2019}. The dynamical age of the jet is also $> 9\times 10^3 \, \rm yr$ \citep{Chema_2012}, which is compatible with both models. Therefore, based on energetic arguments, we cannot discard a hadronic origin for the $\gamma$-ray emission. However, under the IC scattering model, the electron energy required to account for the detected $\gamma$-ray spectrum is $\rm \sim 10^{54} \, erg$, much higher than the other two models. The injection time, considering an efficiency of $\eta = 100\%$, is much longer than the lifetime of the jet. Therefore, IC scattering is expected to be negligible under the jet ambient conditions.

Since the cooling timescales from Fig. \ref{fig:cooling-times} are longer than the jet lifetime, accelerated particles were able to travel $\rm \sim 100 \, kpc$ under a diffusion coefficient of the order of $\rm 10^{28} \, cm^2 s^{-1}$ at particle energies of $\sim 1 \, \rm GeV$ \citep[e.g.][]{dif_coef_97, Voyager_16}. This distance is large enough to cover a significant part of the molecular complex where the source is located, possibly producing weak high-energy emission within an extended area. Thus, we overlay the density maps of HII, HI, and $\rm H_2$ to illustrate the location of the density distribution of the region where the $\gamma$-ray emission is detected.

The HII density is derived from a free-free emission map obtained trough the combined analysis of Plank, WMAP, and 408 MHz Survey data \citep{Planck_HII_survey}. The column density is calculated using the conversion factors from \cite{Finkbeiner_2003} and the Eq. (5) described in \cite{Sodroski_1997}. Similarly, the HI density map in our ROI is obtained from the HI4PI survey \citep{HI4PI_survey}. Additionally, to gain spatial resolution, we derive the H density in the molecular complex L291 using extinction maps based on the reddening of the stars in the field of view \citep{Guver_2009, Bayestar_2019}. We also validated the resulting density values for the molecular cloud with the density of H$_2$ returned by the CO Composite Survey \citep{Dame_2001, Bolatto_2013}.
%Additionally, the density distribution of $\rm H_2$ is derived from the data of the CO Composite Survey \citep{Dame_2001}, employing the CO-to-$\rm H_2$ conversion factor from \cite{Bolatto_2013} ($X_{\rm CO} = 2 \times 10^{20} \rm \, cm^{-2} K^{-1} (km/s)^{-1} $). 

Density maps of HII and HI do not exhibit any spatial correlation with the source. On the other hand, the H density map of L291 seems to trace the $\gamma$-ray emission produced by HH 80-81 of the southern part (see Fig. \ref{fig:mwlMap}). Northern side of the molecular cloud is dominated by two unassociated point-like sources reported in the 4FGL DR4 catalog: J1818.1-2000 ($\rm RA_{J2000}=18^h 18^m 07^s.9, \ Dec_{J2000}= -20^\circ 00' 58''$) and J1817.2-1946c ($\rm RA_{J2000}=18^h 17^m 14^s.9, \ Dec_{J2000}= -19^\circ 46' 48''$). Based on the ATNF Catalog and the IR map in the Fig. \ref{fig:mwlMap}, J1817.2-1946c might be associated to the PSR J1817-1938 pulsar, with only $\sim 8 \, \rm arcmin$ of separation. The global $\gamma$-ray excess provided by the three \textit{Fermi}-LAT sources detected in the molecular cloud is spatially coincident with the densest regions of the molecular cloud, enhancing the idea that the detected $\gamma$-ray excess may have a galactic origin. However, since HH 80-81 is perfectly described as a point-like source and the $\gamma$-ray emission in the area is faint, we cannot discuss further the morphology of the individual source.

\section{Conclusion}
\label{sect:conclusion}

In this work, we present a detailed analysis of the high-energy emission located in the proximities of the protostellar jet of HH 80-81. We analyzed 15 years of \textit{Fermi}-LAT data, improving the exposure time compared to the previous detection by \cite{Yan2022}. Given the faint flux of the $\gamma$-ray excess and its position close to the Galactic plane, we excluded data below 300 MeV, where the Galactic diffuse emission dominates over our target source. In fact, we find discrepancies in the spectral index of the gamma-ray source with respect to the previous study carried out by \cite{Yan2022} between $\rm 100 \, MeV$ and $\rm 300 \, MeV$. With our analysis, we conclude that the low-energy $\gamma$-ray band, combined with shorter exposure times, is severely impacted by the Galactic background, returning inaccurate results.
%In fact, we found significant discrepancies on the $\gamma$-ray spectrum of HH 80-81. The lower energy regime, combined with shorter exposure times, was proved to be unreliable, leading us to conclude that the analysis performed by \cite{Yan2022} is severely impacted due to Galactic template, and their reported steep spectrum is erroneous.

Additionally, we perform a comprehensive source identification based on positional arguments. We find two potential candidates in the region: HH 80-81 protostellar jet, and J1819 compact radio source. We study the possibility that the $\gamma$-ray emission would be originated by both sources. HH 80-81 is a very stable source, with no flares reported so far. In contrast, blazars (as J1819 could be) usually present high variability in the $\gamma$-ray band. Since the $\gamma$-ray source shows non-significant variability and HH 80-81 is the only one counterpart falling inside the $1\sigma$ position area (and also in the $2\sigma$ area), we conclude that HH 80-81 is the main candidate for explaining the detected emission. However, other MYSOs have been detected flaring in IR-optical wavelengths. Thus, a definite proof of the relation between the gamma-ray emission and MYSOs would come from the association of flaring episodes in longer wavelengths with variability in the gamma-ray band. Therefore, IR surveys tracking the flare episodes of these types of objects is crucial to detect more MYSOs emitting $\gamma$ rays as a result of a significant particle acceleration.

The obtained radiative spectrum reveals a harder behaviour than that reported by \cite{Yan2022}, with an updated power-law index of $\Gamma = 2.62 \pm 0.12$ versus a previous value of $3.53 \pm 0.11$. We fit the detected radiation to be originated from either leptonic or hadronic particle distributions. Both populations are compatible with the spectral shape of the source. In addition, the amount of energy injected in both particle distributions is achievable during the jet's lifetime. As a result, the exact nature of the particle population that is producing the $\gamma$-ray emission remains unclear and will require longer exposure times to perform more significant morphological studies. In this way, we display IR and density maps over the detected $\gamma$-ray emission to show the spatial coincidence with the Galactic structures of L291.

%In contrast, we obtained a harder spectrum, with a power-law index of $\Gamma = 2.62 \pm 0.12$, consistent with the spectrum produced by particles being accelerated through DSA. We have elaborated a deep analysis of the $\gamma$-ray production, concluding that both leptonic and hadronic models are compatible with the detected emission in terms of energetic injection. Nevertheless, the clear correlation between the $\gamma$-ray emission and the molecular clouds shown in Fig. \ref{fig:mwlMap} highlights the necessity of further studies to determine whether this correlation has physical significance or is due to an inaccurate modeling of the diffuse background emission. If confirmed, the hadronic model would arise as the primary candidate for explaining the high-energy emission, confirming that protostellar jets are able to accelerate CRs.

%Other MYSOs have been detected flaring in IR-optical wavelengths. However, HH 80-81 is a very stable source, with no flares reported so far. Thus, we rely on positional arguments to correlate the $\gamma$-ray emission with the protostellar object. The possibility of detecting future flares in similar objects will require continuous analysis of their multi-wavelength emission to definitively demonstrate that they are sources of $\gamma$-ray emission, enhancing the conclusions of this paper.

Finally, this work demonstrates that HH objects are excellent candidates for studying the capability of protostellar jets to accelerate particles. Therefore, further investigations of these objects along all the electromagnetic spectrum  are essential to constrain the ambient conditions in order to infer the particle spectrum from the high-energy emission. Regarding HH 80-81, the two main knots (i.e. HH 80 and HH 81) have been studied in radio \citep{Marti_1993, Vig_2018} and X-rays \citep{Pravdo_2004, Rodriguez-Kamenetzky_2019}. Nonetheless, the large PSF of the LAT instrument detects the entire protostellar jet as a point-like source. To combine our findings with multiwavelength data, especially radio and X-ray bands where the non-thermal emission dominates, we need further studies covering the whole region of interest. In the same way, future observations using the Large-Size Telescopes (LSTs) of CTAO could provide key insights into the morphology of the $\gamma$-ray production region, since the theoretical cutoff for the maximum reachable energy of the accelerated particles seems compatible with the minimum energy threshold of the LST telescopes \citep[see][]{Araudo_2021}.

\begin{acknowledgements}
    Authors thank Carlos Carrasco-Gonz\'alez for his useful comments on the source identification. We also thank the comments of the anonymous referee that have undoubtedly improved the quality of this paper.

    EdOW acknowledges the support of DESY (Zeuthen), a member of the Helmholtz Association HGF.
    
    J.M.-G., J.O.-S.,  R.F.,  R.L.-C., acknowledge financial support from the Severo Ochoa grant CEX2021-001131-S funded by MCIN/AEI/ 10.13039/501100011033.

    J.M.-G., J.O.-S.,  R.L.-C also acknowledge financial support from the Spanish "Ministerio de Ciencia e Innovaci\'on" through grant PID2022-139117NB-C44.

    J.M.-G. acknowledges financial support from the FPI-Severo Ochoa grant CEX2021-001131-S-20-6, PRE2022-103386 funded by MICIU/AEI/ 10.13039/501100011033 and ESF+.
    
    J.O.-S. acknowledges financial support from the project ref. AST22\_00001\_9 with founding from the European Union - NextGenerationEU, the "Ministerio de Ciencia, Innovación y Universidades, Plan de Recuperación, Transformación y Resiliencia", the "Consejería de Universidad, Investigación e Innovación" from the "Junta de Andalucía", and the "Consejo Superior de Investigaciones Científicas".
    
    R.F. acknowledges support from the grants Juan de la Cierva FJC2021-046802-I, PID2020-114461GB-I00, and PID2023-146295NB-I00 funded by MCIN/AEI/ 10.13039/501100011033 and by ``European Union NextGenerationEU/PRTR''.

    R.L.-C. acknowledges the Ram\'on y Cajal program through grant RYC-2020-028639-I.

\end{acknowledgements}

\bibliographystyle{aa} % style aa.bst
\bibliography{references} % your references

\end{document}